\documentclass[aps,prl,twocolumn,amsmath,amssymb,nofootinbib,superscriptaddress,floatfix,reprint,longbibliography]{revtex4-1}
\usepackage[dvips]{graphicx}
\usepackage{latexsym}
\usepackage{amsmath}
\usepackage{amsfonts}
\usepackage{amssymb}
\usepackage{bm}
\usepackage{color}
\usepackage{txfonts}
\usepackage{float}
\usepackage{url}
\usepackage[colorlinks=true, urlcolor=blue, linkcolor=blue, citecolor=blue ]{hyperref}
\usepackage{ulem}
\begin{document}
	\newcommand{\fig}[2]{\includegraphics[width=#1]{#2}}
	\newcommand{\la}{{\langle}}
	\newcommand{\ra}{{\rangle}}
	\newcommand{\dg}{{\dagger}}
	\newcommand{\upa}{{\uparrow}}
	\newcommand{\dna}{{\downarrow}}
	\newcommand{\ab}{{\alpha\beta}}
	\newcommand{\ias}{{i\alpha\sigma}}
	\newcommand{\ibs}{{i\beta\sigma}}
	\newcommand{\hH}{\hat{H}}
	\newcommand{\hn}{\hat{n}}
	\newcommand{\hc}{{\hat{\chi}}}
	\newcommand{\hU}{{\hat{U}}}
	\newcommand{\hV}{{\hat{V}}}
	\newcommand{\br}{{\bf r}}
	\newcommand{\bk}{{{\bf k}}}
	\newcommand{\bq}{{{\bf q}}}
	\def\gsim{~\rlap{$>$}{\lower 1.0ex\hbox{$\sim$}}}
	\setlength{\unitlength}{1mm}
	\newcommand{{\vhf}}{$\chi^\text{v}_f$}
	\newcommand{{\vhd}}{$\chi^\text{v}_d$}
	\newcommand{{\vpd}}{$\Delta^\text{v}_d$}
	\newcommand{{\ved}}{$\epsilon^\text{v}_d$}
	\newcommand{{\vved}}{$\varepsilon^\text{v}_d$}
	\newcommand{{\tr}}{{\rm tr}}
	\newcommand{\pprl}{Phys. Rev. Lett. \ }
	\newcommand{\pprb}{Phys. Rev. {B}}

\title {Decoding flat bands from compact localized states}
\author{Yuge Chen}
\thanks{These two authors contributed equally}
\affiliation{Beijing National Laboratory for Condensed Matter Physics and Institute of Physics,
	Chinese Academy of Sciences, Beijing 100190, China}
\author{Juntao Huang}
\thanks{These two authors contributed equally}
\affiliation{Beijing National Laboratory for Condensed Matter Physics and Institute of Physics,
	Chinese Academy of Sciences, Beijing 100190, China}
\affiliation{School of Physical Sciences, University of Chinese Academy of Sciences, Beijing 100190, China}

\author{Kun Jiang}
\email{jiangkun@iphy.ac.cn}
\affiliation{Beijing National Laboratory for Condensed Matter Physics and Institute of Physics,
	Chinese Academy of Sciences, Beijing 100190, China}
\affiliation{School of Physical Sciences, University of Chinese Academy of Sciences, Beijing 100190, China}

\author{Jiangping Hu}
\email{jphu@iphy.ac.cn}
\affiliation{Beijing National Laboratory for Condensed Matter Physics and Institute of Physics,
	Chinese Academy of Sciences, Beijing 100190, China}
\affiliation{Kavli Institute of Theoretical Sciences, University of Chinese Academy of Sciences,
	Beijing, 100190, China}

\date{\today}

\begin{abstract}
The flat band system is an ideal quantum platform to investigate the kaleidoscope created by the electron-electron correlation effects. The central ingredient of realizing a flat band is to find its compact localized states. In this work, we develop a systematic way to generate the compact localized states by designing destructive interference pattern from 1-dimensional chains. A variety of 2-dimensional new flat band systems are constructed with this method.  Furthermore, we show that the method  can be extended to generate the compact localized states  in  multi-orbital systems by carefully designing the block hopping scheme, as well as in  quasicrystal and disorder systems.
\end{abstract}
\maketitle

The flat band with a dispersionless band structure is one special system with intriguing effects in condensed matter \cite{flatband_review,flatband_review2,yang19,Sutherland86,Mielke1,Mielke2,Mielke3}. This dispersionless feature ensures its zero group velocity and infinite effective mass quasiparticles, which also induces a divergent density of states (DOS). Therefore, the external responses of this ``super heavy'' system are dramatically enriched, like the divergent orbital magnetic susceptibility \cite{quantum_distance} and the negative orbital magnetism found in kagome magnet Co$_3$Sn$_2$S$_2$ \cite{jiaxin19}. More interestingly, the interplay between correlation and the flat band leads to various quantum many-body phenomena including ferromagnetism, Wigner crystal formation, fractional quantum hall states, etc. \cite{fm1,fm2,wucj07,frqh1,frqh2,frqh4,frqh3}. Recently, the emergence of twistronics in bilayer graphene and transition metal dichalcogenides also makes the flat band investigation into a new stage with correlation, superconductivity, and topology \cite{tbg_review,tbg1,tbg2,tbg3}.
Hence, how to find more flat band systems and how to understand their flatness are now becoming alluring questions in quantum matter.
Historically, different methods have been applied to finding flat bands including line graphs, compact localized states, bipartite sublattices etc. \cite{Mielke1,Mielke2,Mielke3,Bernevig_n,Bernevig_np,CLS0,CLS1,CLS2,CLS3,Aoki96}. But how to quickly locate the flat band remains a question.

\begin{figure}
\centering
  \includegraphics[width=8.5cm]{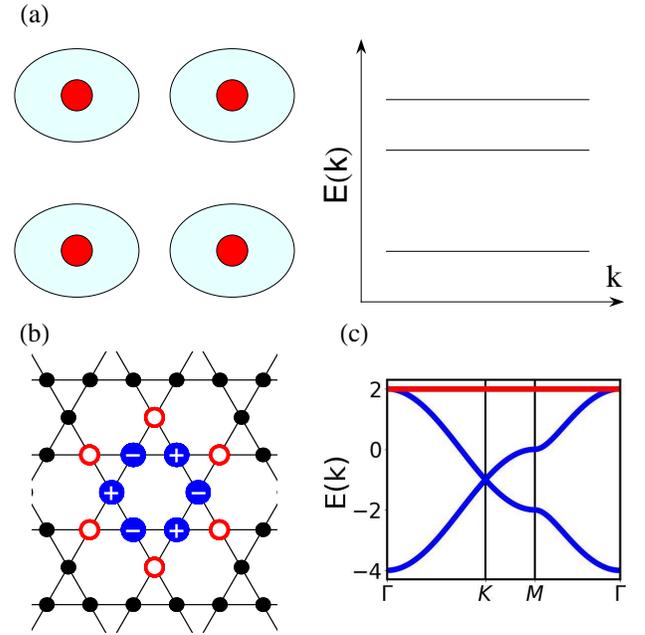}\\
  \caption{(a) The atomic limit of one solid without hybridization between each identical atom. The band structure of this system is strictly flat at each eigenenergy shown on the left. (b) A compact localize state in kagome lattices. Red open circles correspond to the wavefunction zero of the compact localized state. The positive (+) and negative (-) signs correspond to the compact localized state wavefunction weight at each blue site.
  (c) The energy dispersion of kagome lattice, where the flat band is highlighted by the red line.}\label{kagomeCLS}
\end{figure}

Generally speaking, a flat band is similar to the atomic limit of the band structure in solid-state physics. As illustrated in Fig.\ref{kagomeCLS}(a), if we arrange many identical atoms into a solid lattice without hybridization, the energy levels of each atom are the exact same with translation symmetry. The band structures of this atomic limit are their flat atomic energy levels as in Fig.\ref{kagomeCLS}(a). Therefore, the dispersionless nature of the flat band also implies its spacial localization feature. This localized orbital is one special localized eigenstate of the system called the compact localized state \cite{CLS0,CLS1,CLS2,CLS3,Sutherland86,Aoki96,yang19}, whose amplitude is finite only inside a confined region in real space, and vanishes outside of it.
The kagome lattice is a typical example of the flat band system, as shown in Fig.\ref{kagomeCLS}(b). The compact localized state of the kagome lattice is localized inside its hexagonal plaquette with sign alternative wavefunction weight, as highlighted in Fig.\ref{kagomeCLS}(b). Since neighbor sites have opposite weights, the particle cannot hop outside its hexagon due to \textit{destructive interference} resulting in the flat band in Fig.\ref{kagomeCLS}(c).
Hence, one direct approach toward flat band systems is finding compact localized states. In this work, we develop a systematic way of generating compact localized states and constructing flat bands in 2-dimensional lattices utilizing compact localized states. We also generalize the above results into the multi-orbital systems, quasicrystal, and disorder systems.

More specifically, when the Hamiltonian of a crystal exists an eigenstate completely localized in a finite region, we call it a compact localized state. Because of the crystal translational symmetry, a translated localized state remains the eigenstate of the Hamiltonian. These energy degenerate compact localized states lead to the flat band in momentum space after Fourier transformation. Since the wavefunction outside the localized region is strictly zero,
the arrangement of the lattice outside this region does not affect the compact localized state as the eigenstate of the total Hamiltonian. Therefore, finding a crystal system with localized states is equivalent to finding a finite-size system with an eigenstate with wavefunction vanishing on its boundary. The ``star of David" compact localized state in the kagome lattice is one standard example.

\begin{figure}
\centering
  \includegraphics[width=7.5cm]{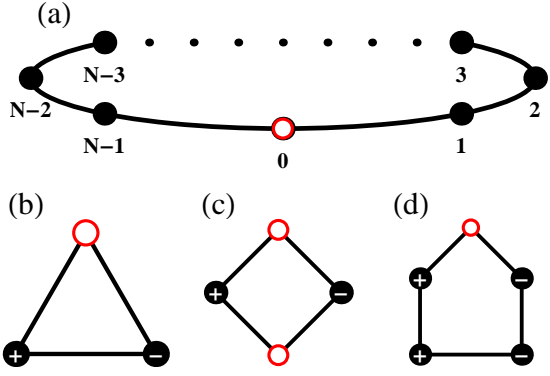}\\
  \caption{(a) A 1D closed-loop chain with $N$ sites. (b) $N=3$ chain and its eigenstates with zeros. The red lattice indicates the wavefunction weight is zero. The $\pm$ signs indicate the weight sign at each site. The destructive interference is the essential reason for the wavefunction node. (c) $N=4$ chain and its eigenstates with zeros. (d) $N=5$ chain and its eigenstates with zeros.}\label{1dc}
\end{figure}
In order to find the compact localized states, we start from the closed loop 1-dimensional (1D) chain with $N$ sites, as illustrated in Fig.\ref{1dc} (a). The Hamiltonian of this chain can be written as:
\begin{equation}
H_{1D}=t \sum _{i=1}^N \left(C_i^{\dagger }C_{i+1}+C_{i+1}^{\dagger }C_i\right),
\end{equation}
where $t$ is hopping strength and set to be 1. To simplify our discussion, we will only consider the nearest neighbor hoppings with the same amplitude in the following discussion.
By Fourier transformation, the Hamiltonian becomes
\begin{equation}
H_{1D}=\sum _{k} E(k) C_k^{\dagger }C_k
\end{equation}
where $E(k)=2t\cos k$. The eigenstate can be easily obtained as $\psi _{k}=\frac{1}{\sqrt{N}}\sum _{i=1}^N e^{i k r_i}C_i$. By noticing that $E(k)=E(-k)$ for a general $k$, we can always construct a new eigenstate:
\begin{equation}
\psi=\psi _{k}-\psi _{-k}=\frac{1}{N}\sum _{i=1}^N \left(e^{i k r_i}-e^{-i k r_i}\right)C_i
\end{equation}
Clearly, the wavefunction of $\psi$ vanishes at the origin $r_i=0$. Then, we arrive at the building blocks with zeros ($red$ circle sites in graphs) shown in Fig.\ref{1dc} (b-d) with $N=3,4,5$ respectively. And one can generate any finite numbers $\psi$ with zeros at the origin.
Notice that the wavefunctions in Fig.\ref{1dc} (b-d) contain similar \textit{destructive interference} with sign alternative weights connecting the $red$ sites in Fig.\ref{kagomeCLS}(b).

\begin{figure}
\centering
  \includegraphics[width=7.5cm]{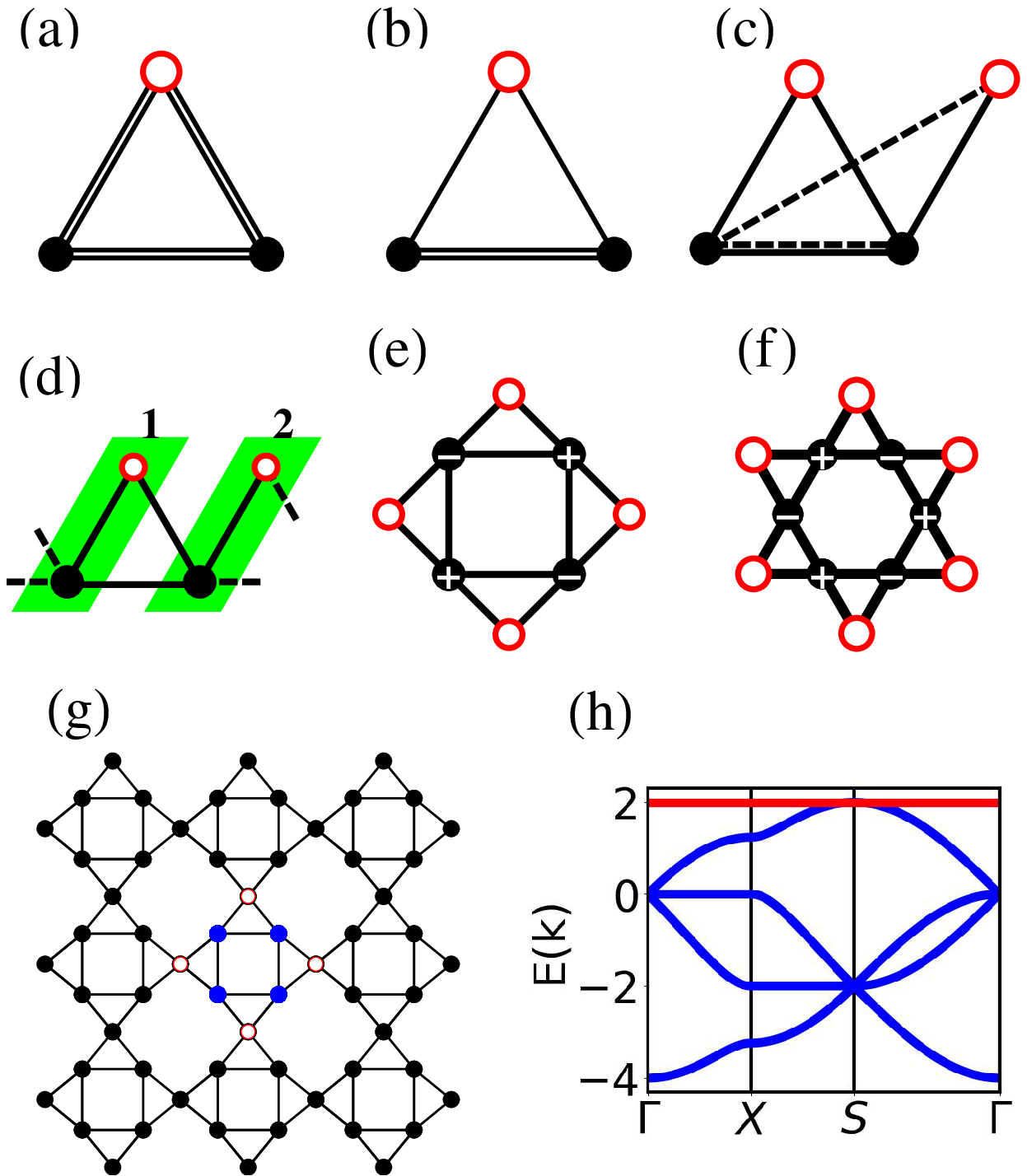}\\
  \caption{(a-d) The constructing procedures of 2D compact localized states from the $N=3$ chain. The red lattice in the graph represents the one eigenstate with zeros on the red lattices in those systems. Both the solid lines and the dashed lines represent a hopping between the two lattices, and the number of lines between the two lattices (solid or dashed) represents the relative strength of the bond. (e) The square compact localized state from $N=3$ chain. (f) The kagome compact localized state from $N=3$ chain. (g) The 2D flat band lattice from gluing the compact localized states in (e). (h) The band structure of (g) obtained from PythTB.
  }\label{3LC}
\end{figure}

After obtaining $\psi$, the next step is to build a finite system with wavefunction zeros at its boundary. The basic idea is to keep the \textit{destructive interference} pattern by linking the building blocks in Fig.\ref{1dc}. We first take $N=3$ as an example.
Graphically, we insert a new $red$ site into Fig.\ref{3LC}(a) and rescale the hopping strength between $red$ sites with black sites, as shown in Fig.\ref{3LC}(c). We can prove that this new state is also an eigenstate with zeros at $red$ sites.


The proof contains two steps. The first statement we want to prove is the eigenstate $\psi$ remains eigenstate after one can rescale the hopping strength between red sites and other sites, as shown in Fig.\ref{3LC}(b). The $\psi$ state is the eigenstate with energy $E$ can be written as
\begin{eqnarray}
H\psi=\left(\begin{array}{cc}
 H_{b} & T  \\
 T^{\dagger } & \epsilon_0 \\
\end{array}\right)\left(\begin{array}{c}
 \psi_b  \\
 0 \\
\end{array}\right)=E\left(\begin{array}{c}
 \psi_b  \\
 0 \\
\end{array}\right)
\end{eqnarray}
where $H_{b}$ is the Hamiltonian without $red$ cites, $\epsilon_0$ is the onsite energy of red sites and $T$, $T^\dagger$ are the hopping matrix between $red$ sites and black sites. The $\psi_b$ is the $\psi$ wavefunction without zeros. Rescaling the $T$ by a factor $c$, we can prove that $\psi$ remains an eigenstate with energy $E$ as
\begin{eqnarray}
\left(\begin{array}{cc}
 H_{b} & cT  \\
 cT^{\dagger } & \epsilon_0 \\
\end{array}\right)\left(\begin{array}{c}
 \psi_b  \\
 0 \\
\end{array}\right)=E\left(\begin{array}{c}
 \psi_b  \\
 0 \\
\end{array}\right)
\end{eqnarray}
Furthermore, we can couple another red site into the system as Fig.\ref{3LC}(c).
The new $\psi$ with an additional zero in the new red site remains an eigenstate with energy $E$ because the following equation remains valid :
\begin{equation}
\left(\begin{array}{ccc}
 H_{b} & cT  & cT  \\
 cT ^{\dagger } & \epsilon_0  & T' \\
 cT ^{\dagger } & T'^{\dagger} & \epsilon_0'  \\
\end{array}\right)\left(\begin{array}{c}
 \psi_b  \\
 0 \\
 0 \\
\end{array}\right)=E\left(\begin{array}{c}
 \psi_b  \\
 0 \\
 0 \\
\end{array}\right)
\end{equation}
where $T'$ is the coupling between two red sites and $\epsilon_0$, $\epsilon_0'$ are their onsite energy respectively.

Following the above two steps, we prove that this new state is still an eigenstate in Fig.\ref{3LC}(c) and the graph method is one efficient
way of generating eigenstates with zeros. Using the graph method, we redraw Fig.\ref{3LC}(c) by connecting black sites periodically as shown in Fig.\ref{3LC}(d). Furthermore, we can copy and link the graph Fig.\ref{3LC}(d) and arrive at the graph Fig.\ref{3LC}(e).
This new state is also an eigenstate keeping the \textit{destructive interference}. To prove this, we split the $\psi$ in Fig.\ref{3LC}(d) graph into two parts $\psi_1$, $\psi_2$. The eigenstate equation becomes
\begin{eqnarray}
\left(
\begin{array}{cc}
 H_{11} & T_{12}+T_{21}^{\dagger } \\
 T_{21}+T_{12}^{\dagger } & H_{22} \\
\end{array}
\right)\left(
\begin{array}{c}
 \psi _1 \\
 \psi _2 \\
\end{array}
\right)=E\left(
\begin{array}{c}
 \psi _1 \\
 \psi _2 \\
\end{array}
\right)
\end{eqnarray}
where $H_{11/22}$ are the matrices within $\psi_1$/$\psi_2$. $T_{12}$ is the hopping matrix between $\psi_1$,$\psi_2$ (solid links) while $T_{21}$ is the hopping matrix between $\psi_2$,$\psi_1$ (dash links).
Then, we can find a new eigenstate satisfying the following eigenequation
\begin{eqnarray}
\left(\begin{array}{cccc}
 H_{11} & T_{12} & 0 & T_{21}^{\dagger } \\
 T_{12}^{\dagger } & H_{22} & T_{21} & 0 \\
 0 & T_{21}^{\dagger } & H_{11} & T_{12} \\
 T_{21} & 0 & T_{12}^{\dagger } & H_{22} \\
\end{array}\right)\left(
\begin{array}{c}
 \psi _1 \\
 \psi _2 \\
 \psi _1 \\
 \psi _2 \\
\end{array}
\right)=E\left(
\begin{array}{c}
 \psi _1 \\
 \psi _2 \\
 \psi _1 \\
 \psi _2 \\
\end{array}\right)
\end{eqnarray}
This eigenstate is exactly the wavefucntion in Fig.\ref{3LC}(e). In a similar way, one can further generate graph Fig.\ref{3LC}(f), which is indeed the kagome compact localized state discussed above. Now we have two finite-size systems containing eigenstates with wavefunction zeros along their boundary. By duplicating these two graphs and keeping the translation symmetry, we can construct the flat band system efficiently.
For example, we can link the graph Fig.\ref{3LC}(e) periodically at their zero corners and obtain lattice Fig.\ref{3LC}(g). And this lattice contains a flat band in its dispersion, as shown in Fig.\ref{3LC}(h). The band dispersions are calculated based on PythTB package \cite{pythontb}.

Here, we still want to emphasize that the compact localized state is not the Wannier function \cite{flatband_review,flatband_review2}. They do not need to form an orthonormal set and cover more than one unit cell. For example, the compact localized state in the kagome lattice (Fig.\ref{kagomeCLS}) covers two unit cells while Fig.\ref{3LC}(f) compact localized state is within one unit cell. There are two types of flat band systems: an orthonormal set of compact localized states with vanishing overlap and
a non-orthonormal set of compact localized states with finite overlap.

\begin{figure}
\centering
  \includegraphics[width=7.5cm]{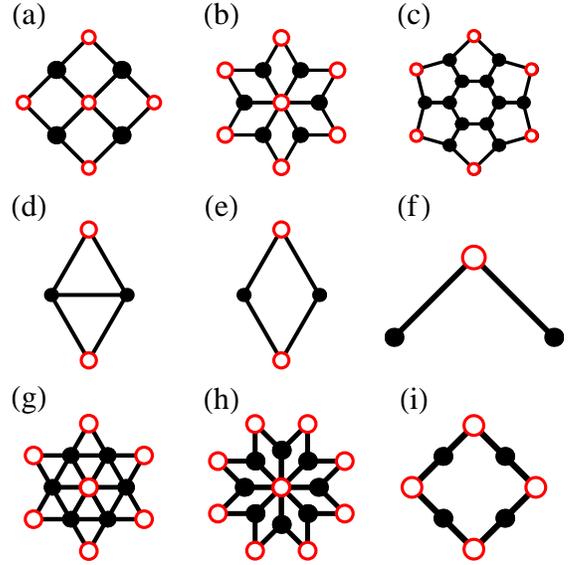}\\
  \caption{Additional examples of compact localize state from $N=3,4,5$ chains. The labeling convention is the same as above. These figures provide the ``Lego" construction sets for flat band systems discussed below.}\label{cls}
\end{figure}

Following the same procedures in $N=3$ patterns, we can use Fig.\ref{1dc} (c-d) with $N=4$ and $N=5$ patterns to generate compact localized states. By reserving the \textit{destructive interference} feature, we get the graphs in Fig.\ref{cls} (a-c). Fig.\ref{cls} (a-b) are coming from $N=4$ pattern while Fig. \ref{cls} (c) is from $N=5$ pattern. The detailed construction process can be found in the supplemental material \cite{sm}.
Furthermore, there are also two additional graphs Fig.\ref{cls} (d-e) by rescaling the hopping strength between Fig.\ref{3LC} (c). We can also cut the link between black sites Fig.\ref{1dc} (b) arriving graph Fig.\ref{cls} (f).
Then, we easily generate the compact localized states in Fig.\ref{cls} (g-i) using Fig.\ref{cls} (d-f).

\begin{figure*}
	\begin{center}
		\fig{7.0in}{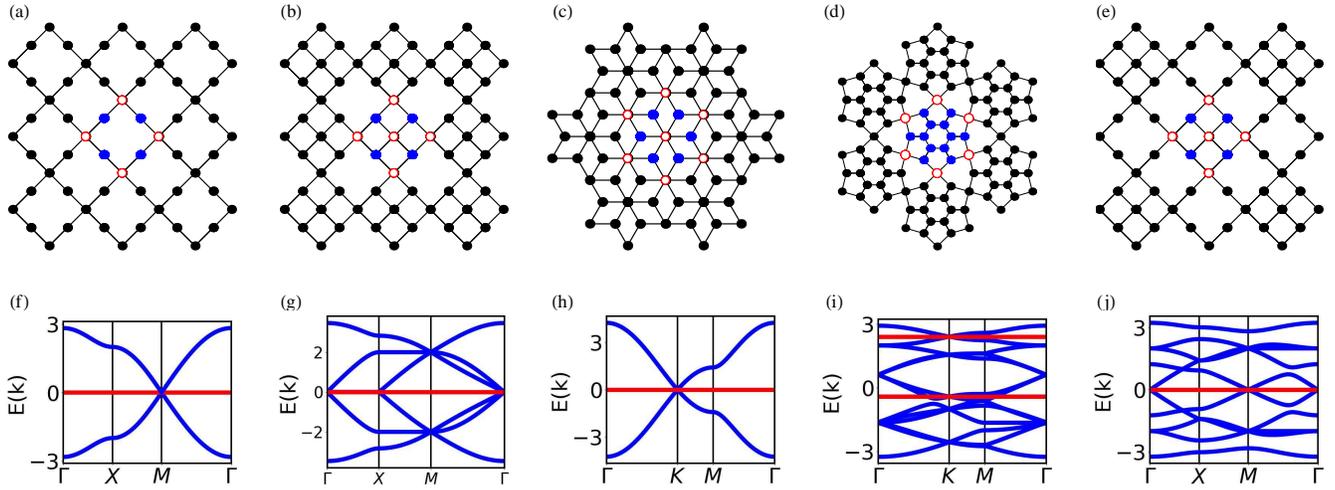}
		\caption{Examples of the 2D flat band systems constructed from the above compact localized states in Fig.\ref{3LC} and Fig.\ref{cls}. Their compact localized states are highlighted with the same labeling convention. (f-j) are their corresponding band structures. (a) is the Lieb lattice. (c) is the Dice lattice.}\label{example}
	\end{center}
\end{figure*}

After obtaining all these compact localized states, we can treat them like the ``Lego" construction sets for flat bands. Carefully gluing these construction sets and reserving the translation symmetry, we successfully build various 2-dimensional flat band systems as shown in Fig.\ref{example} (a-e). Their corresponding band structures are plotted in Fig.\ref{example} (f-j). Fig.\ref{example} (a) is the well-known flat band Lieb lattice. Fig.\ref{example} (c) is another well-known flat band Dice lattice. All of the above examples contain flat bands as the red highlighted lines in Fig.\ref{example} (f-j). Therefore, our method is one quick and direct way of obtaining flat band systems. Furthermore,
we also list other flat band examples with their compact localized states in Fig.\ref{example2}. Their corresponding band structures can be found in the supplemental materials \cite{sm}.

\begin{figure*}
	\begin{center}
		\fig{7.0in}{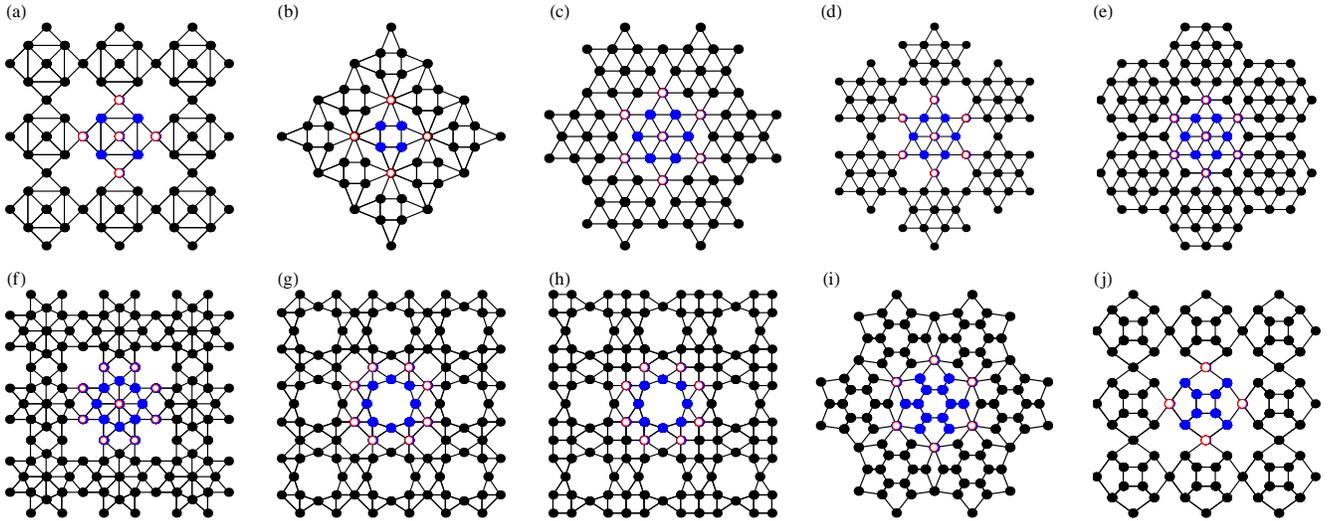}
		\caption{ Additional examples of 2D flat band systems and their compact localized states. The labeling convention is the same as above.}\label{example2}
	\end{center}
\end{figure*}

The \textit{destructive interference} plays an essential role in constructing the above compact localized states in single-orbital models. Naturally, another question arises: whether we can generalize this scheme to multi-orbital systems. In realistic materials, the complex multi-orbital situation is more common than the above single-orbital model. Hence, finding multi-orbital flat band systems may provide a new way of realizing orbit physics in condensed matter.
The multi-orbital flat band system has already been achieved in the degenerate $p_x$-$p_y$ model on a honeycomb lattice \cite{wucj07}. And this system is supposed to be realized in high-precision optical lattices \cite{xuzhifang,Lixiaopeng}. To simplify our discussion, we will focus on the nearest neighbor two orbital $p_x$-$p_y$ model here and leave other multi-orbital systems for future discussion.

\begin{figure}
	\begin{center}
		\fig{3.4in}{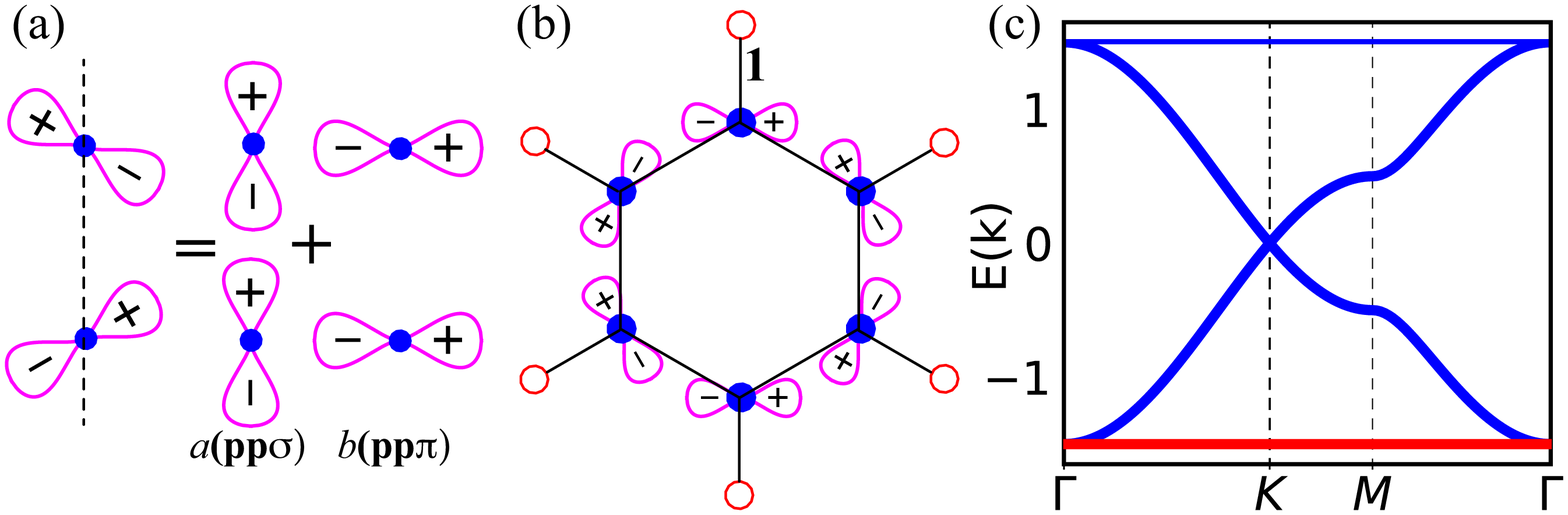}
		\caption{(a) Two $p$ orbital hopping along $y$ direction can be decomposed into the linear combination $a(pp\sigma)$ and $b(pp\pi)$, where $a=\frac{\sqrt{3}}{2}$ and $b=-\frac{\sqrt{3}}{2}$ in this particular cases. (b) The compact localized state in $p_x$-$p_y$ honeycomb lattice. This state is localized inside the hexagon owing to the corresponding $p$ orbitals being perpendicular to their escape bonds.
  (c) The band structure with the red flat band of $p_x$-$p_y$ honeycomb lattice model. The hopping parameter here is $(pp\sigma)=1$ and $(pp\pi)=0$.
  }\label{px-py}
	\end{center}
\end{figure}

Owing to their special atomic wavefunction geometry, the hopping integrals between orbitals are highly anisotropic.
A commonly used strategy is the linear combination of atomic orbitals (LCAO), which can be simplified by the Slater-Koster parameters \cite{marder2010condensed,slater-koster}. There are only two Slater-Koster parameters $(pp\sigma)$ and $(pp\pi)$ for $p$ orbitals.
In the language of chemistry, the $(pp\sigma)$ and $(pp\pi)$ are the $\sigma$ and $\pi$ chemical bonds between $p$ orbitals respectively.
As illustrated in Fig.\ref{px-py} (a), the hopping between any two $p$ orbitals can be decomposed into the linear combination of $(pp\sigma)$
and $(pp\pi)$ while other combinations are zero due to symmetry.

We can always tune the ratio between $(pp\sigma)$ and $(pp\pi)$ to achieve the same compact localized states through the above \textit{destructive interference} scheme. The resulting two-orbital flat bands also become double degenerate. Since these compact localized states are similar to those above, we leave this part of the discussion in the supplemental materials \cite{sm}.

\begin{figure}
	\begin{center}
		\fig{3.4in}{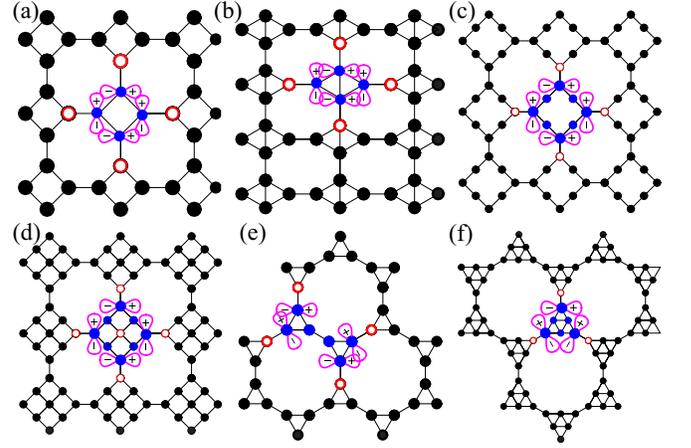}
		\caption{Examples of $p_x$-$p_y$ flat band systems in 2D and their corresponding compact localized states. The orbital patterns for each state are highlighted. The hopping bond outside each state is blocked due to vanishing $(pp\pi)$.}\label{px-py-lattice}
	\end{center}
\end{figure}

Interestingly, there is also another compact localized state scheme in the multi-orbital system beyond \textit{destructive interference}.
The \textit{destructive interference} utilizes the interference between two bonds.
This multi-orbital scheme is achieved by blocking one bond hopping. Thus, we can name it the \textit{block hopping} scheme.
This phenomenon has been found in the honeycomb lattice $p_x$-$p_y$ model \cite{wucj07}.
Generally speaking, $(pp\pi)$ is weaker than $(pp\sigma)$. We can take $(pp\pi)$ zero.
As shown in Fig.\ref{px-py}(b), a compact localized state centered at the hexagonal plaquette is constructed in the honeycomb lattice.
At each plaquette site, the corresponding $p$ orbital is perpendicular to its bond outside this hexagon.
Take bond-1 in Fig.\ref{px-py} (b) as one example, bond-1 is linked to a $p_x$ orbital along the $y$ direction. The only hope for a $p_x$ orbital hops along $y$ bond is through the vanishing $(pp\pi)$ bond as illustrated in Fig.\ref{px-py} (a).
Hence, the hopping along bond-1 is blocked for this state resulting in a compact localized state. The corresponding $p_x$-$p_y$ band structure in the honeycomb lattice contains two flat bands as shown in Fig.\ref{px-py} (c).
Using this \textit{block hopping} idea, we can easily generalize into other lattices by constructing the p-orbital compact localized states. As plotted in Fig.\ref{px-py-lattice}, we obtain six p-orbital flat band systems with their highlighted corresponding compact localized states.
Each bond hopping outside the compact localized states is blocked as in the honeycomb lattice, as highlighted in Fig.\ref{px-py-lattice}. Their corresponding band structures can be found in the supplemental materials \cite{sm}.

Finally, we want to generalize the above ideas to the quasicrystals, even the disordered systems. Although there are no translation periodicity and bands in quasicrystals and disordered cases, the localized eigenstates remain valid features. These localized eigenstates can also lead to divergent DOS and corresponding conventional responses under correlations or external fields.

\begin{figure}
	\begin{center}
		\fig{3.4in}{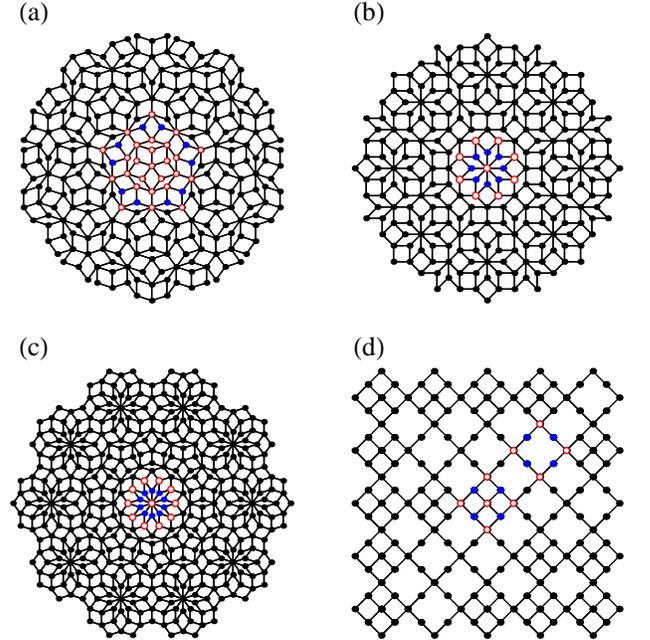}
		\caption{Quasicrystals and disorder systems with highlighted localized states. (a) The Penrose lattice with 5-fold rotation symmetry. (b) the quasicrystal lattice with 8-fold rotation symmetry. (c) The quasicrystal lattice with 10-fold rotation symmetry.
  (d) A disordered system with randomly distributing compact localized states in Fig.\ref{3LC}(d) and Fig.\ref{cls}(e).}\label{qc}
	\end{center}
\end{figure}

As shown in Fig.\ref{qc}(a), the Penrose lattice is a classical model of quasicrystals with 5-fold rotational symmetry. The Penrose lattice indeed contains the compact localized states, as highlighted in Fig.\ref{qc}(a). This feature has been pointed out many years ago by Semba and Ninomiya in the center model and by Kanhmoto and Sutherland in the vertex model \cite{quasi-crystal-5b,quasi-crystal-5a}. Constructing this compact localized state is also straightforward as we discussed above. Beyond 5-fold quasicrystal, there are also quasi-crystal models with 8-fold rotational symmetry \cite{quasi-crystal-8} and 10-fold rotational symmetry. We illustrate two quasicrystal examples and their compact localized states in Fig.\ref{qc}(b-c). Furthermore, the disordered systems by randomly distributing compact localized states can be constructed as plotted in Fig.\ref{qc}(d).


In summary, we develop a systematic way of generating compact localized states.
Using the finite-size 1-dimensional chain, we can always find wavefunction with zeros at the origin. By gluing the 1-D chain and keeping the \textit{destructive interference}, we can quickly construct various compact localized states. Then, the various 2D flat band systems are found by linking these compact localized states through translation operators.
We cathato generalize this idea to multi-orbital systems. Besides the \textit{destructive interference}, there is another \textit{block  hopping} scheme for the multi-orbital cases. Using the $p_x$-$p_y$ model as an example, we succeed in constructing various multi-orbital flat band systems.
In the end, we apply the above compact localized states into quasi-crystal systems and disorder systems.
We hope that these findings could provide a new understanding of flat band and stimulate the search for other flat band systems.

\textit{Acknowledgement}
This work is supported by the Ministry of Science and Technology  (Grant No. 2022YFA1403901), the National Natural Science Foundation of China (Grant No. NSFC-11888101, No. NSFC-12174428), the Strategic Priority Research Program of the Chinese Academy of Sciences (Grant No. XDB28000000), and the Chinese Academy of Sciences through the Youth Innovation Promotion Association (Grant No. 2022YSBR-048).

\bibliography{reference}

\clearpage
\onecolumngrid
\begin{center}
\textbf{\large Supplemental Material: Decoding flat bands from compact localized states}
\end{center}

\setcounter{equation}{0}
\setcounter{figure}{0}
\setcounter{table}{0}
\setcounter{page}{1}
\makeatletter
\renewcommand{\theequation}{S\arabic{equation}}
\renewcommand{\thefigure}{S\arabic{figure}}
\renewcommand{\thetable}{S\arabic{table}}
\renewcommand{\bibnumfmt}[1]{[S#1]}
\renewcommand{\citenumfont}[1]{S#1}

\twocolumngrid

\subsection{Create 2D flat band systems by 1D chains (N = 4, 5)}
We have introduced the procedures of getting compact localized states by the 1D chain with N = 3. But for the situations of N = 4 and N = 5, each has one more procedure to make the compact localized states more reasonable.

\begin{figure}
\centering
 \includegraphics[width=8.5cm]{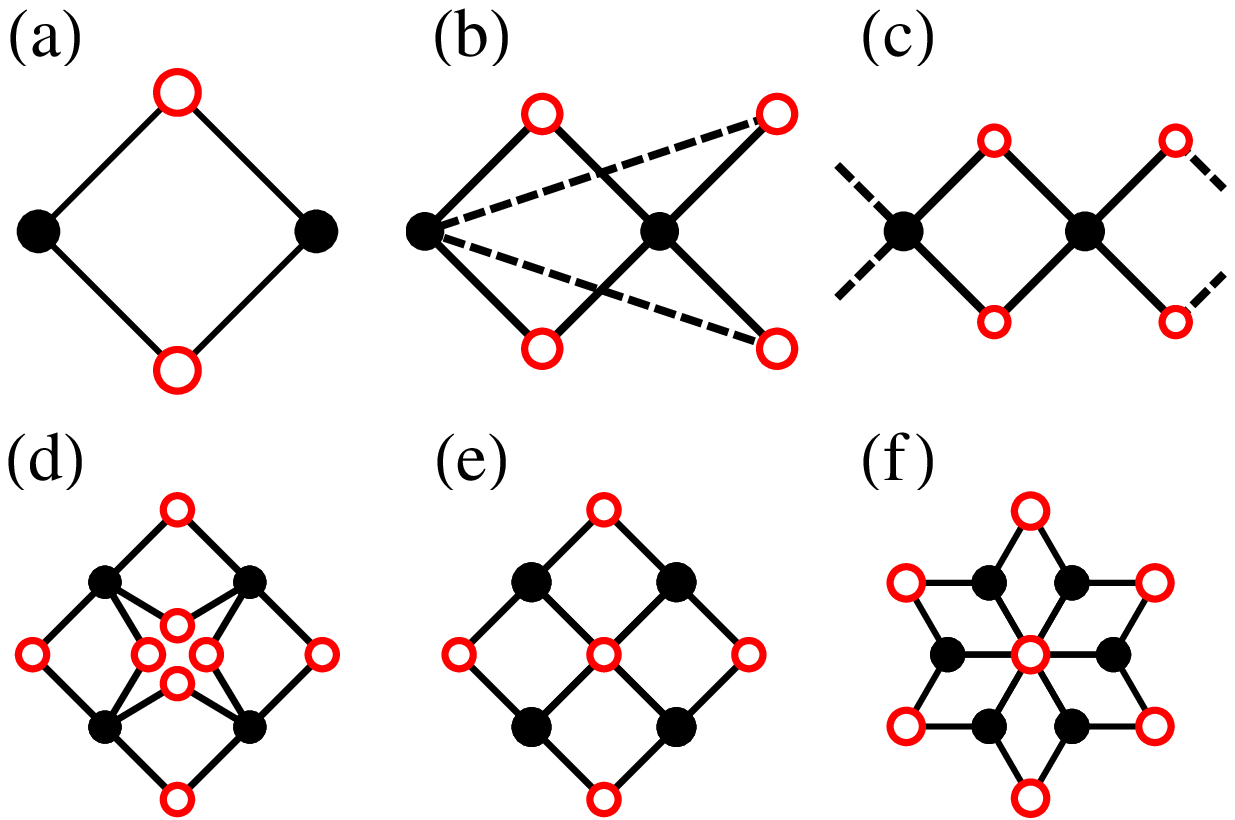}\\
 \caption{(a-c) The constructing procedures of 2D compact localized states from the N = 4 chain. The red lattice in the graph represents the one eigenstate with zeros on the red lattices in those systems. Both the solid lines and the dashed lines represent a hopping between the two lattices, and the number of lines between the two lattices (solid or dashed) represents the relative strength of the bond. (d) The compact localized state before the fusing step. (e-f) The compact localized states from the N = 4 chain after the fusing step.}\label{4LC}
\end{figure}

For the chain with N = 4, after the same procedures as N = 3, we will get the graph Fig.\ref{4LC}(d). But we want there to be only one lattice in the center of the graph as we showed in Fig.\ref{4LC}(e) and Fig.\ref{4LC}(f) rather than four lattices. So we fuse the center lattices of the graph Fig.\ref{4LC}(d) one by one to get the graph Fig.\ref{4LC}(e) and the same for graph Fig.\ref{4LC}(f).

 For prove fusing step, we assume there is a system with an eigenstate which has two zero wavefunction lattices and with energy E, which can be written as:
\begin{equation}
\left(
\begin{array}{ccc}
 H_1 & \alpha  & \beta  \\
 \alpha ^{\dagger } & a & b \\
 \beta ^{\dagger } & b* & c \\
\end{array}
\right)\left(
\begin{array}{c}
 \psi  \\
 0 \\
 0 \\
\end{array}
\right)=E\left(
\begin{array}{c}
 \psi  \\
 0 \\
 0 \\
\end{array}
\right)
\end{equation}
where $H_1$ is the Hamiltonian of the system without any zero wavefunction lattice. The vector $\psi$ is the wave function on $black$ lattices, the vectors $\alpha$, $\beta$, $\alpha^\dagger$, and $\beta^\dagger$ are the hopping from one zero wavefunction lattice to others and others to that zero wavefunction lattice. $a$ and $b$ are the onsite energy of two zero wavefunction lattices, and $c$ are the hopping strength between the two zero wavefunction lattices.

Fusing those two lattices, the new state $\phi$ with one less zero remains an eigenstate with energy E because the following equation remains valid:
\begin{equation}
\left(
\begin{array}{cc}
 H_{1} & \alpha + \beta  \\
 \alpha ^{\dagger } + \beta ^{\dagger } & f \\
\end{array}
\right)\left(
\begin{array}{c}
 \psi  \\
 0 \\
\end{array}
\right)=E\left(
\begin{array}{c}
 \psi  \\
 0 \\
\end{array}
\right)
\end{equation}
where $f$ is the new onsite energy of the fused point. After a few fusing steps and one more rescaling step of the hopping strength, we can get graph Fig.\ref{4LC}(e) from Fig.\ref{4LC}(d). In the same way, we can also get graph Fig.\ref{4LC}(f).
\begin{figure}
\centering
  \includegraphics[width=8.5cm]{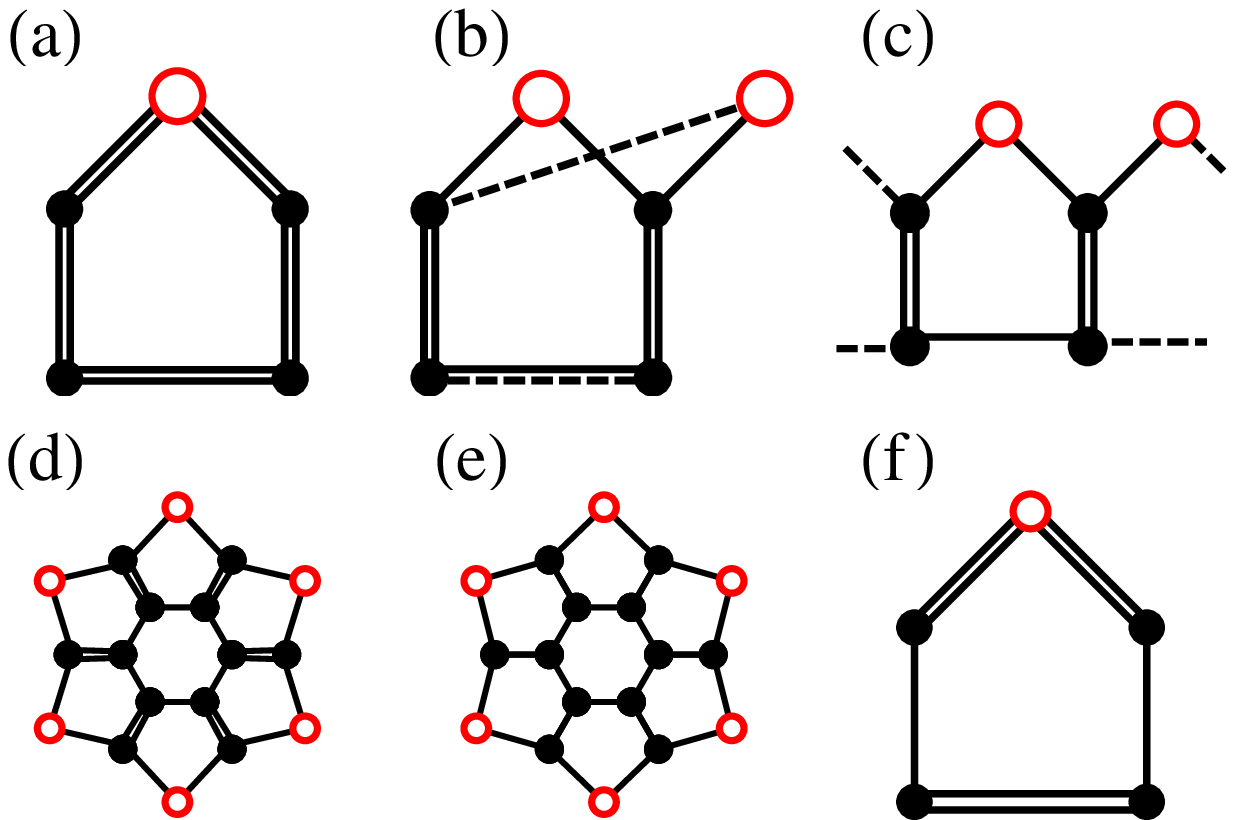}\\
  \caption{(a-c) The constructing procedures of 2D compact localized states from the N = 5 chain. The labeling convention is the same as above. (d) The compact localized state before rescaling the hopping strength step. (e) The compact localized states from the N = 5 chain after rescaling the hopping strength step in the beginning. (f) The rescaled hopping strength chain.}\label{5LC}
\end{figure}

But for the chain with N = 5, the procedure we want to add is not at the end but at the beginning. That is because if we follow the same procedures as N = 3, we will get the graph Fig.\ref{5LC}(d) from Fig.\ref{5LC}(a). But if we want the hopping strength in the graph Fig.\ref{5LC}(d) to be unified as the graph Fig.\ref{5LC}(e). We need to add a procedure before the beginning to change the hopping strength of the chain. Fig.\ref{5LC}(f) is the system we got after changing the hopping strength of Fig.\ref{5LC}(a). Because of the mirror symmetry, wavefunction zero is still there. Then we can easily get the graph Fig.\ref{5LC}(e) from the graph Fig.\ref{5LC}(f) by the same method.

\begin{figure*}
	\begin{center}
		\fig{7.0in}{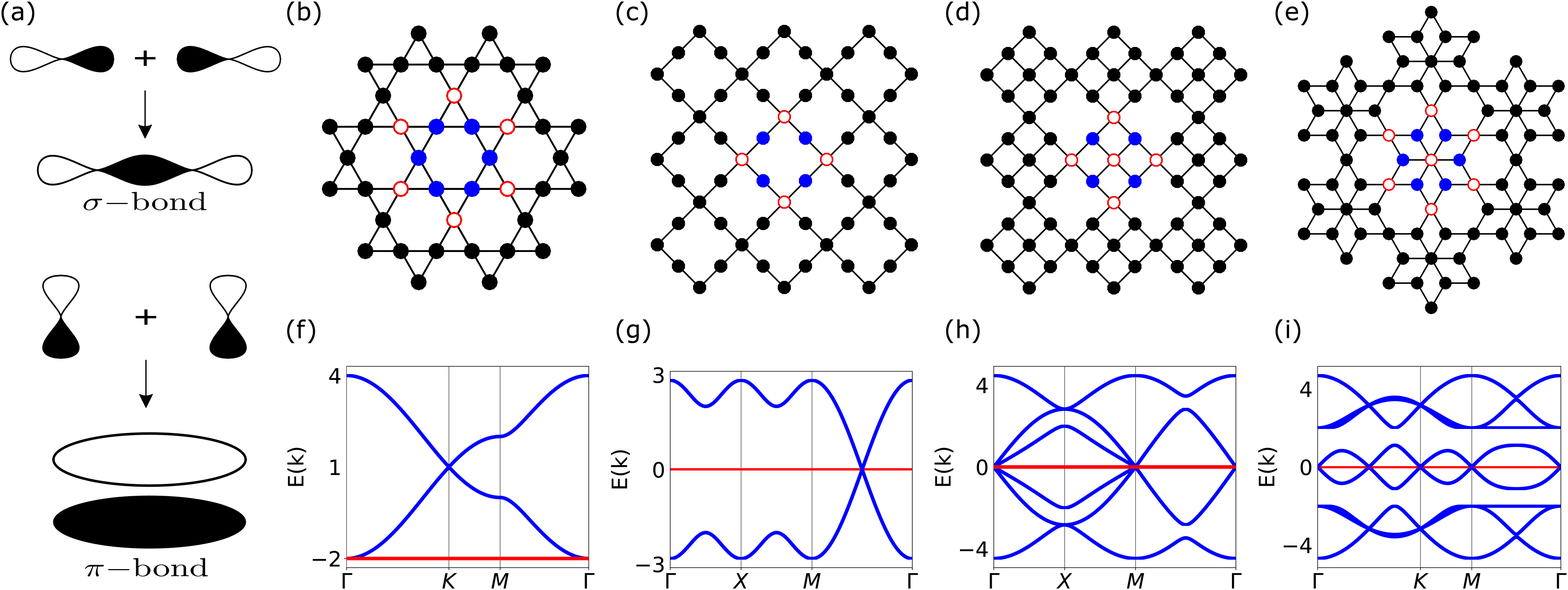}
		\caption{(a) Diagrammatic sketch of $\pi-$bond and $\sigma-$bond for $p_{x}-p_{y}$ orbitals. $(pp\sigma)$ is denoted by $t_{1}$ and $(pp\pi)$ is denoted by $t_{2}$. (b)-(e) Examples of 2D flat band systems in the $p_{x}-p_{y}$ model with $t_{2}=t_{1}$ and their compact localized states. (f)-(i) Corresponding energy bands for the 2D flat band systems in (b)-(e) with the flat band indicated by the red line.  }\label{example2-1}
	\end{center}
\end{figure*}

\subsection{Multi-orbital systems from single-orbital systems}
Fig.\ref{example2-1}(a) is a schematic diagram of $(pp\sigma)$ bond and $(pp\pi)$ bond. For the $\sigma$ type bond, the hopping between p orbitals is parallel to the orientation along the bond while for the $\pi$ type bond the hopping between p orbitals is perpendicular to the bond direction. We set $t_{1}=(pp\sigma), t_{2}=(pp\pi)$ in the $p_{x}-p_{y}$ model. Fig.\ref{example2-1}(b)-(e)  are examples of two dimensional flat band systems in the $p_{x}-p_{y}$ model with $t_{2}=t_{1}$. When $t_{1}=t_{2}$, the hopping matrix between the sites at the nearest neighbor atoms is proportional to the identity matrix by the LCAO method. Therefore, the compact localized states of the two-orbit model are obtained by the direct product of the compact localized states of the corresponding single-orbit model and the 2D basis vectors representing $\{p_{x},p_{y}\}$ orbital degree. Their corresponding energy bands are plotted in Fig.\ref{example2-1}(f)-(i) with the flat band marked as red lines. Fig.\ref{example2-1}(b) is the well-known kagome lattice. Its band structure in $p_{x}-p_{y}$ model(Fig.\ref{example2-1}(f)) is just a two-fold degenerate analog of the single-orbit counterpart. Fig.\ref{example2-1}(c) is the famous Lieb lattice in the two-orbital model. These results in $p_{x}-p_{y}$ orbitals show the same compact localized states as that of the single-orbit model when $t_{2}=t_{1}$.

\subsection{Band structures of the systems in Fig.\ref{example2} and Fig.\ref{px-py-lattice} }
\begin{figure*}
	\begin{center}
		\fig{7.0in}{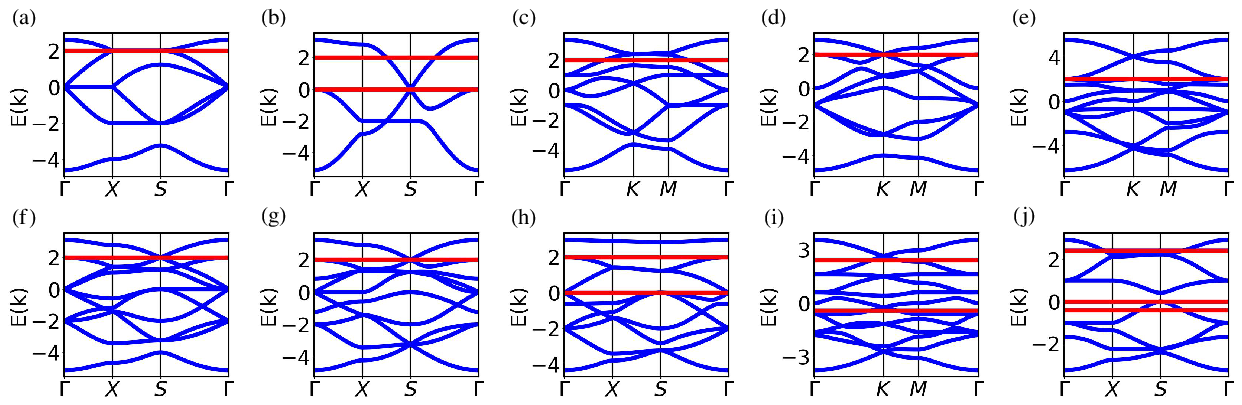}
		\caption{Band structures of the systems in Fig.\ref{example2} (in main texture). Flat bands are highlighted by the red lines. }\label{SEk}
	\end{center}
\end{figure*}

In the main text, we have discussed various flat band examples in Fig.\ref{example2}. Their corresponding band structures are plotted in Fig.\ref{SEk}.

\begin{figure*}
	\begin{center}
		\fig{7.0in}{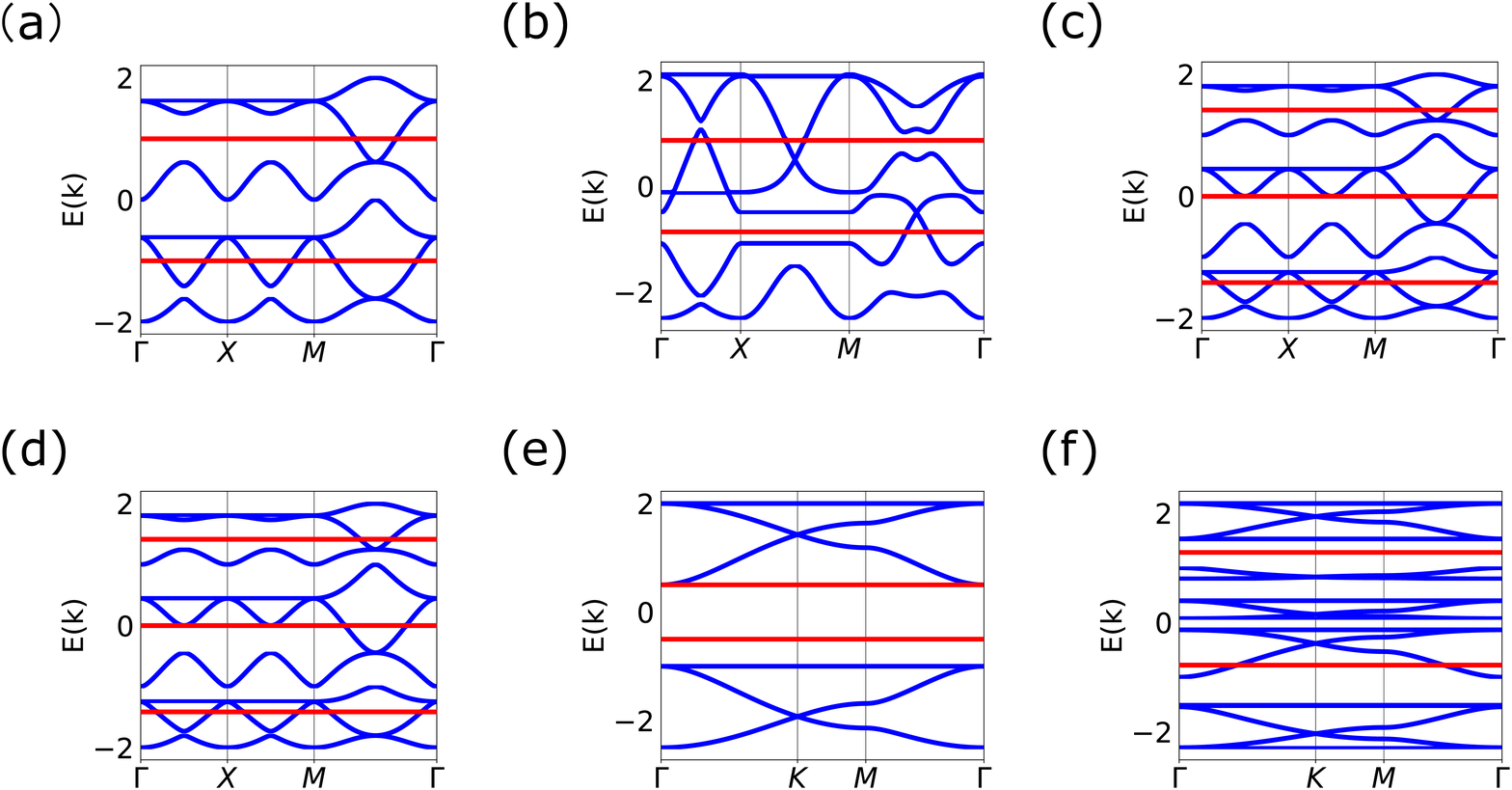}
		\caption{(a)-(f) Energy bands corresponding to 2D flat band systems in (a)-(f) in Fig.\ref{px-py-lattice}. Red lines are the flat bands corresponding to the compact localized states.     }\label{example2-0}
	\end{center}
\end{figure*}

Also in the main text, we have discussed various flat band examples in Fig.\ref{px-py-lattice}. Their corresponding band structures are plotted in Fig.\ref{example2-0}.

\end{document}